\documentclass[a4paper,11pt]{article}
\usepackage{pos}
\usepackage[separate-uncertainty = true,multi-part-units=single]{siunitx}
\DeclareSIUnit\clight{\text{\ensuremath{c}}}
\usepackage{subcaption}
\usepackage{cleveref}
\usepackage{comment}

\title{Progress on the MALTA CMOS Sensor: Edge-TCT Measurements for Parametric Pixel Simulation}
\ShortTitle{Edge-TCT Measurements for Parametric Pixel Simulation}

\author*[a,b]{L. Fasselt}
\author[c]{P. Behera}%\email{}
\author[a]{D. V. Berlea}%\email{}
\author[d]{D. Bortoletto}%\email{}
\author[e]{C. Buttar}%\email{} Craig
\author[c]{T. Chembakan}%\email{}
\author[f]{V. Dao}%\email{} Valerio
\author[c]{G. Dash}%\email{}
\author[g]{Y. Enari}
\author[h]{S. Haberl}%\email{}
\author[h]{T. Inada}%\email{}
\author[i]{F.K. Isik}%\email{}
\author[a,b]{C. Issever}
\author[j]{K. Krizka}
\author[j]{L. Li}%\email{}
\author[k]{X. Li}%\email{}
\author[h]{H. Pernegger}%\email{}
\author[h]{P. Riedler}%\email{}
\author[h]{W. Snoeys}%\email{}
\author[h]{C.A. Solans Sánchez}%\email{}
\author[h]{A. Swoboda}%\email{}
\author[i]{I. Turk Cakir}%\email{}
\author[h]{M. van Rijnbach}%\email{}
\author[c]{A. Vijay}%\email{}
\author[a,b]{S. Worm}%\email{}
\affiliation[a]{Deutsches Elektronen-Synchrotron DESY, Zeuthen, Germany}
\affiliation[b]{Humboldt University of Berlin, Berlin, Germany}
\affiliation[c]{Indian Institute of Technology Madras, Chennai, India}
\affiliation[d]{University of Oxford, Oxford, UK}
\affiliation[e]{University of Glasgow, Glasgow, UK}
\affiliation[f]{Stony Brook University, New York, US}
\affiliation[g]{KEK, Tsukuba, Japan}
\affiliation[h]{CERN, Geneva, Switzerland}
\affiliation[i]{Ankara University, Ankara, Turkey} %
\affiliation[j]{University of Birmingham, Birmingham, UK} %\orgdiv{School of Physics and Astronomy},
\affiliation[k]{Los Alamos National Laboratory, Los Alamos, US} %
\renewcommand{\logo}{\relax} % for arxiv preprint without PoS logo

\emailAdd{lucian.fasselt@desy.de}

\abstract{
MALTA2 is a monolithic active pixel sensor, developed in the \SI{180}{\nano\meter} CMOS imaging technology for radiation-hard applications in high-rate environments. 
This contribution summarizes recent developments in the characterization and simulation of MALTA2. 
The charge collection of a sensor on epitaxial silicon is studied with the Edge Transient Current Technique (Edge-TCT), using a digital readout, where the charge is reconstructed solely from binary hit data.
In previous works, only specially designed analog output pixels at the matrix periphery were investigated.
In contrast, the digital readout gives access to the full edge of the pixel matrix.
Laser scans across neighboring pixels enable a two-dimensional parameterization of the charge collection.
This parameterization serves as input to a parametric pixel simulation, which has been developed for chip design optimization and system-level studies. 
The simulation results are validated with test beam data.
}

\FullConference{
34th International Workshop on Vertex Detectors (VERTEX2026)\\
20-24 July 2026\\ 
Stoos, Switzerland
}

\begin{document}
\maketitle

\section{Introduction}
Future tracking detectors in high-energy physics demand improvements in sensor technologies.
Monolithic active pixel sensors (MAPS) have already demonstrated such improvements over hybrid detectors in terms of lower material budget, higher granularity and lower power consumption.
However, further development is needed for MAPS application in environments with high hit rates and radiation levels~\cite{Snoeys_MAPS}.
The MALTA2 sensor addresses both requirements.
Its design in the \SI{180}{\nano\meter} Tower CMOS imaging process features a small $\mathrm{n}$-type collection electrode in each pixel for low capacitance and noise.
An asynchronous readout architecture provides a time resolution of \SI{2}{\nano\second} for hit rates up to $\SI{100}{\mega\hertz}/\si{\square\centi\meter}$~\cite{RadiationHard}.
On a Czochralski substrate, the sensor is fully efficient up to fluences of $3\times10^{15}\,\SI{1}{\mega\electronvolt}\,\mathrm{n_{eq}}/\si{\square\centi\meter}$.
Increased doping concentrations are being investigated to compensate for irradiation effects.
%Process modifications are being investigated to increase the radiation hardness further.

The optimization of sensors and detectors requires a simulation that is fast and realistic.
This work follows a data-driven approach, that does not rely on proprietary information or assumptions about the fabrication, such as doping profiles and implant sizes.
%, while also keeping the computation time short.
Instead, the charge collection efficiency of MALTA2 is measured directly and modeled for use in simulation.
For this purpose, the sensor edge is scanned with an infrared laser using the Edge transient current technique (Edge-TCT).
%The measurement combines digital readout with  approach of Edge-TCT with digital readout

\Cref{sec:DigETCT} presents the setup and the charge reconstruction from digital hit data.
From these measurements, a model for the charge collection is derived in \cref{sec:1DCCModel}.
The model serves as input to the parametric simulation summarized in \cref{sec:ParSim}, which is validated with test beam data.

\section{Digital Edge-TCT}
\label{sec:DigETCT}

\begin{figure} 
\centering
\begin{subfigure}[t]{0.3\textwidth} 
\centering
\includegraphics[width=\linewidth]{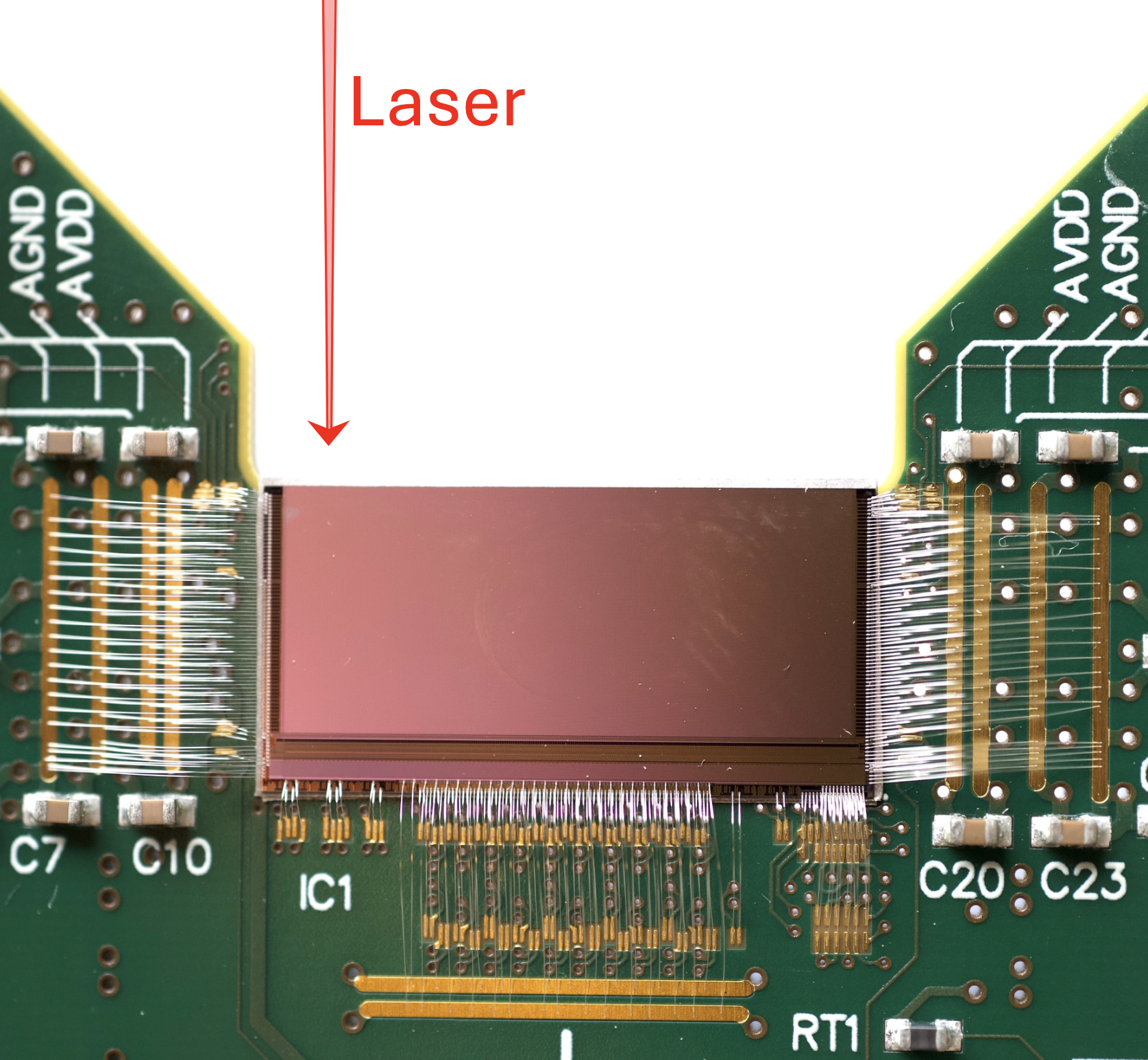}
\caption{Photo}\label{fig:ETCTPhoto}
\end{subfigure}
\hfill
\begin{subfigure}[t]{0.28\textwidth}
\centering
\includegraphics[trim=0 5 0 0,clip,width=\linewidth]{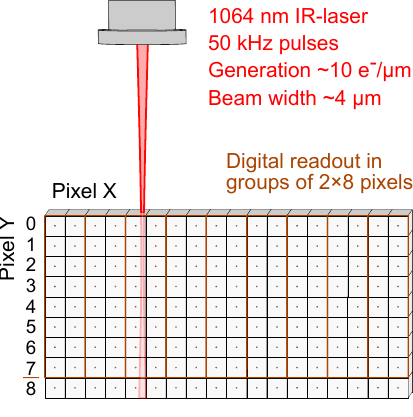}
\caption{Illustration}\label{fig:ETCTSetup}
\end{subfigure}
\hfill
\begin{subfigure}[t]{0.4\textwidth}
\centering
\includegraphics[trim=15 9 5 12,clip,width=\linewidth]{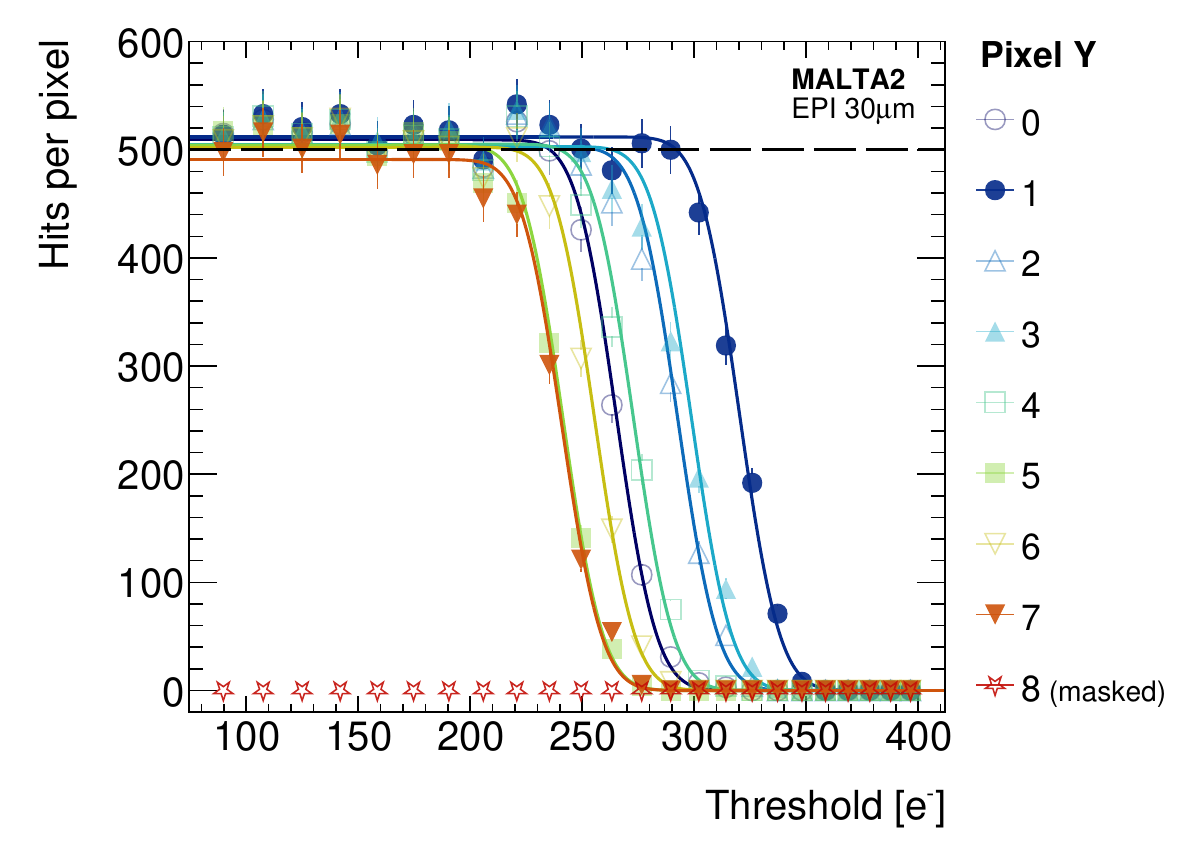}
\caption{Digital data}\label{fig:ThreshScan}
\end{subfigure}
\caption{(a) Photo of a MALTA2 sensor wire-bonded to a PCB designed for Edge-TCT.
The \SI{1064}{\nano\meter} laser is focused on the polished sensor edge as illustrated in (b). 
(c) shows the hits as a function of threshold from the top eight pixels of the same column ($\mathrm{Y}=0$ to 7).
All other rows are masked.
}\label{fig:ETCTIntro}
\end{figure}

A MALTA2 sensor is wire-bonded to a dedicated PCB for Edge-TCT, as shown in \cref{fig:ETCTPhoto}.
It allows scanning of the polished sensor edge~\cite{Berlea_TCT}.
The sensor has a \SI{30}{\micro\meter} thick epitaxial layer grown on a $\mathrm{p}$-type substrate and each pixel contains a front-end that digitizes a hit above threshold.
%The front-end of each pixel contains a discriminator whose threshold is set globally.
TCT measurements with such a digital readout have previously been demonstrated for other MAPS that are illuminated through the backside~\cite{ARCADIA_TCT}.
Here, as illustrated in \cref{fig:ETCTSetup}, an infrared laser ($\lambda = \SI{1064}{\nano\meter}$) is focused through the edge and induces a signal in multiple pixels along the beam (Y-direction).
Its wavelength corresponds to an absorption length of \SI{1}{\milli\meter} in silicon at room temperature~\cite{SiAbsorptionL}.
The laser intensity is therefore approximately constant within a single pixel pitch of \SI{36.4}{\micro\meter}.
The laser is set to deposit about $350\,\mathrm{e}^-$ per pixel, corresponding to about $10\,\mathrm{e}^-$ per \si{\micro\meter}, a factor of six lower than for a minimum-ionizing particle.
This charge lies within the threshold range of 100 to $400\,\mathrm{e}^-$ at the standard gain setting, which is necessary for the charge reconstruction.
%Analog Edge-TCT scans have investigated the amplitude of a monitoring pixel of MALTA2~\cite{Berlea_TCT}.

\subsection{Charge reconstruction from digital data}\label{sec:ChargeReco}
The charge is reconstructed by measuring the number of hits as a function of the discriminator threshold, as shown in \cref{fig:ThreshScan} for pixels located in the same column.
The top eight pixels of the column ($0 \leq Y \leq 7$) are scanned simultaneously, while all other pixels ($\mathrm{Y}\geq8$) are masked to reduce the data rate.
Each threshold is scanned for \SI{10}{\milli\second} at a laser pulse frequency of \SI{50}{\kilo\hertz}.
At low thresholds, all 500 pulses are registered.
Once the threshold exceeds the signal amplitude, the number of hits decreases.
The amplitude of each pixel is therefore reconstructed as the 50\% occupancy point of an error-function fit.
%The uncertainty in the reconstructed amplitude is dominated by the 3\% uncertainty of the charge calibration~\cite{MALTA2Calibration}.
The amplitude generally decreases with the Y-coordinate, because the laser is absorbed in the silicon along the beam.
An exception is the first row ($\mathrm{Y}=0$), in which the pixels show a systematically lower amplitude in all scans.
These pixels have a reduced charge collection due to their location near the edge of the sensor.
The second row of pixels ($\mathrm{Y}=1$) features the largest amplitude and is assumed to be representative of the pixel matrix.
This row is selected for the following scans.

\subsection{Two-dimensional scan}

\begin{figure} 
\centering
\begin{subfigure}[t]{0.44\textwidth} 
\centering
\includegraphics[width=\linewidth]{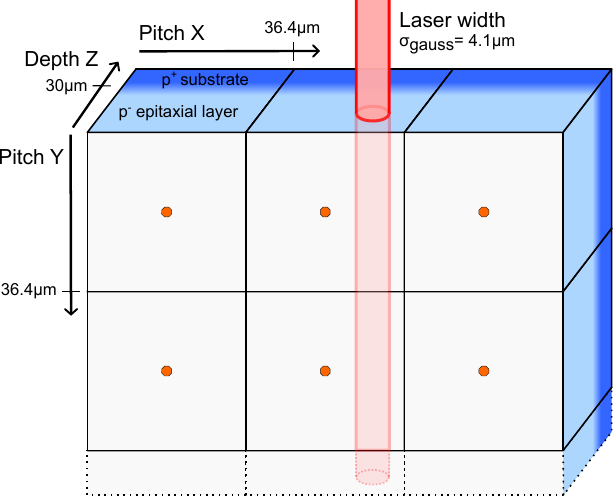}
\caption{Edge scan of a $3\times2$ pixel matrix.}\label{fig:2DDrawing}
\end{subfigure}
\hfill
\begin{subfigure}[t]{0.53\textwidth}
\centering
\includegraphics[width=\linewidth]{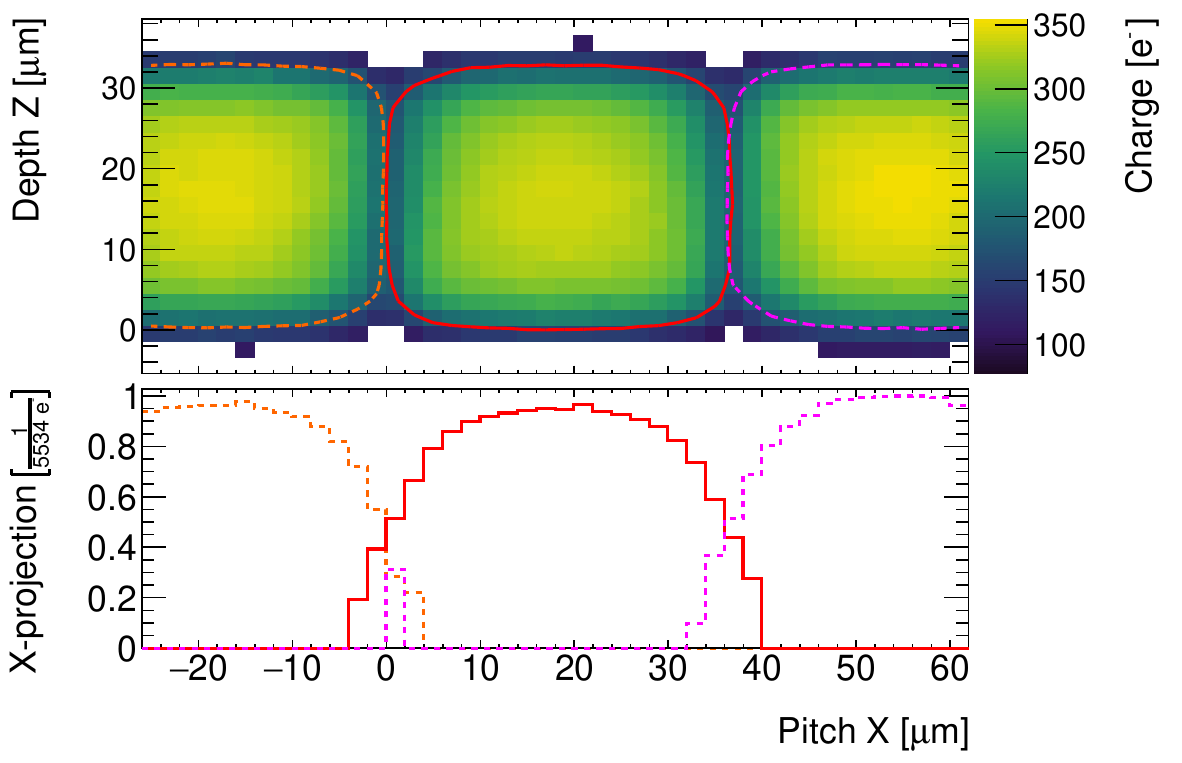}
\caption{Reconstructed charge distribution of three pixels.}\label{fig:2DData}
\end{subfigure}
\caption{Illustration (a) and measurement (b) of a laser scanning the sensor edge.
The charge is reconstructed from digital data and shows the collection profile.
The contours represent the 50\% level of each pixel relative to its maximum charge.
The individual pixel contributions are shown in the X-axis projection.
}\label{fig:2DResults}
\end{figure}

\Cref{fig:2DDrawing} illustrates a 2D scan across the sensor edge (X-Z plane).
The laser is moved in \SI{2}{\micro\meter} steps and the collected charge is reconstructed at each step.
The result is shown in \cref{fig:2DData} for three adjacent pixels in the row $\mathrm{Y=1}$.%, where each bin shows the highest charge of the three pixels.
The high charge regions of the individual pixels are distinguished in the epitaxial layer and reach a maximum charge of $350\,\mathrm{e}^-$.
For each pixel, a contour line marks the interpolated 50\% level relative to its maximum charge.
These contours coincide at the pixel boundaries, as expected from equal charge sharing without significant charge losses.
The equal charge sharing is supported by the projections onto the X-axis in the bottom panel, which intersect at the pixel boundaries at 50\%.
Such a direct measurement of charge sharing is performed for the first time for MALTA2.
The digital readout permits the simultaneous measurement of neighboring pixels, whereas an analog test pixel can only be studied individually.

%While analog test pixels exist, they are all located in the same column (Y) on the side of the sensor and thus do not give access to the measurement of neighbors in the row dimension (X).
The bottom panel of \cref{fig:2DData} shows a feature of the digital readout at $\mathrm{X}=\SI{0}{\micro\meter}$, the boundary of two pixels.
A reconstructed charge appears in the third pixel (dashed violet line), far outside its geometric region.
This is attributed to the digital merging of hits that arrive within a \SI{1.6}{\nano\second} time window.
The merging can shift a hit to another pixel coordinate.
The merging effect as well as its design optimization is discussed in detail in ref.~\cite{FastSim}.
Merging is mitigated in the design of MALTA3, the next sensor generation.
Since the displaced hits lie outside the geometric region of the pixel, they are excluded from the charge collection study.

\begin{figure} 
\centering
\begin{subfigure}[t]{0.47\textwidth}
\centering
\includegraphics[width=\linewidth]{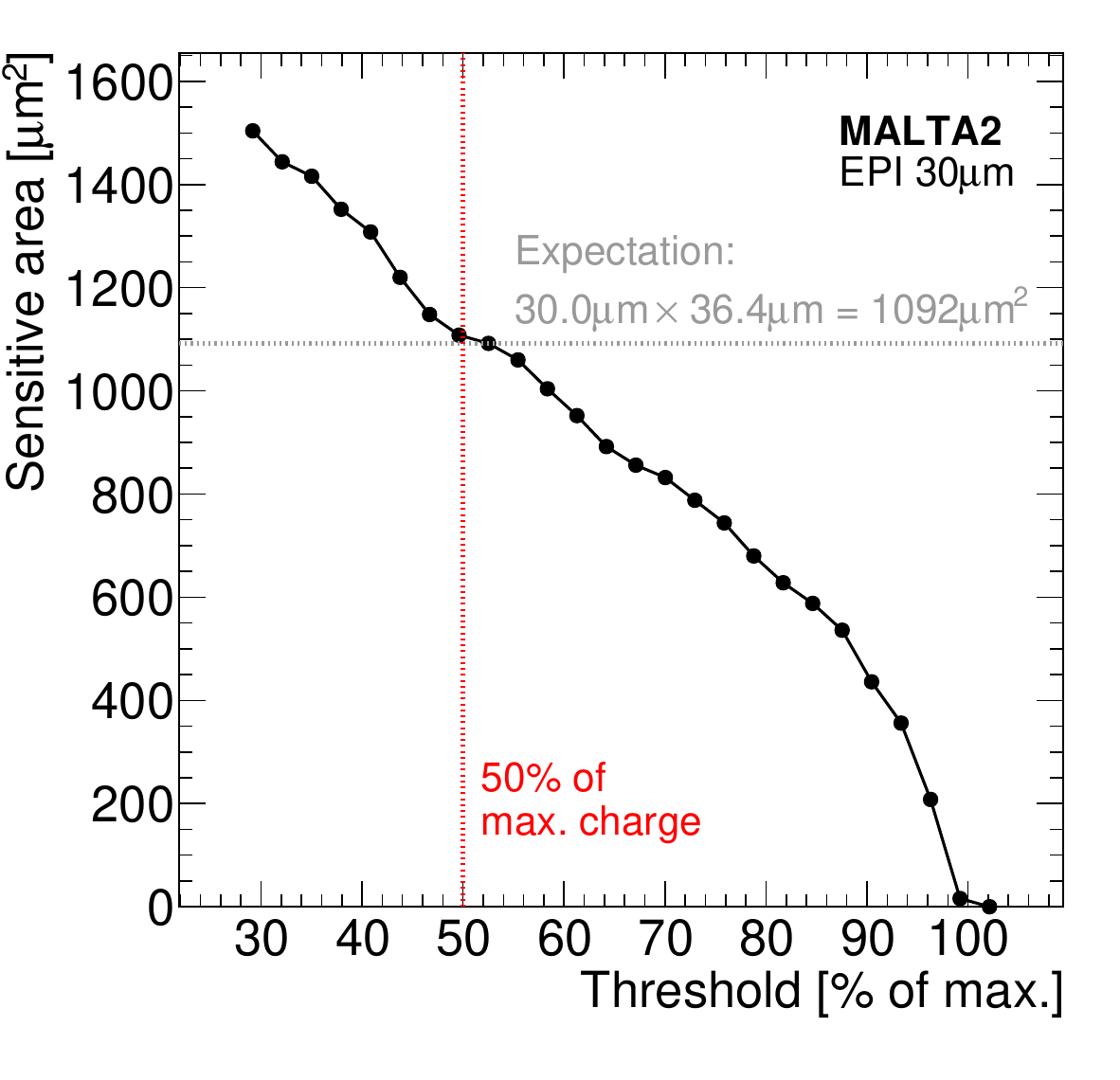}
\caption{Sensitive area of a pixel versus detection threshold.
}\label{fig:SensitiveArea}
\end{subfigure}
\hfill
\begin{subfigure}[t]{0.47\textwidth} 
\centering
\includegraphics[width=\linewidth]{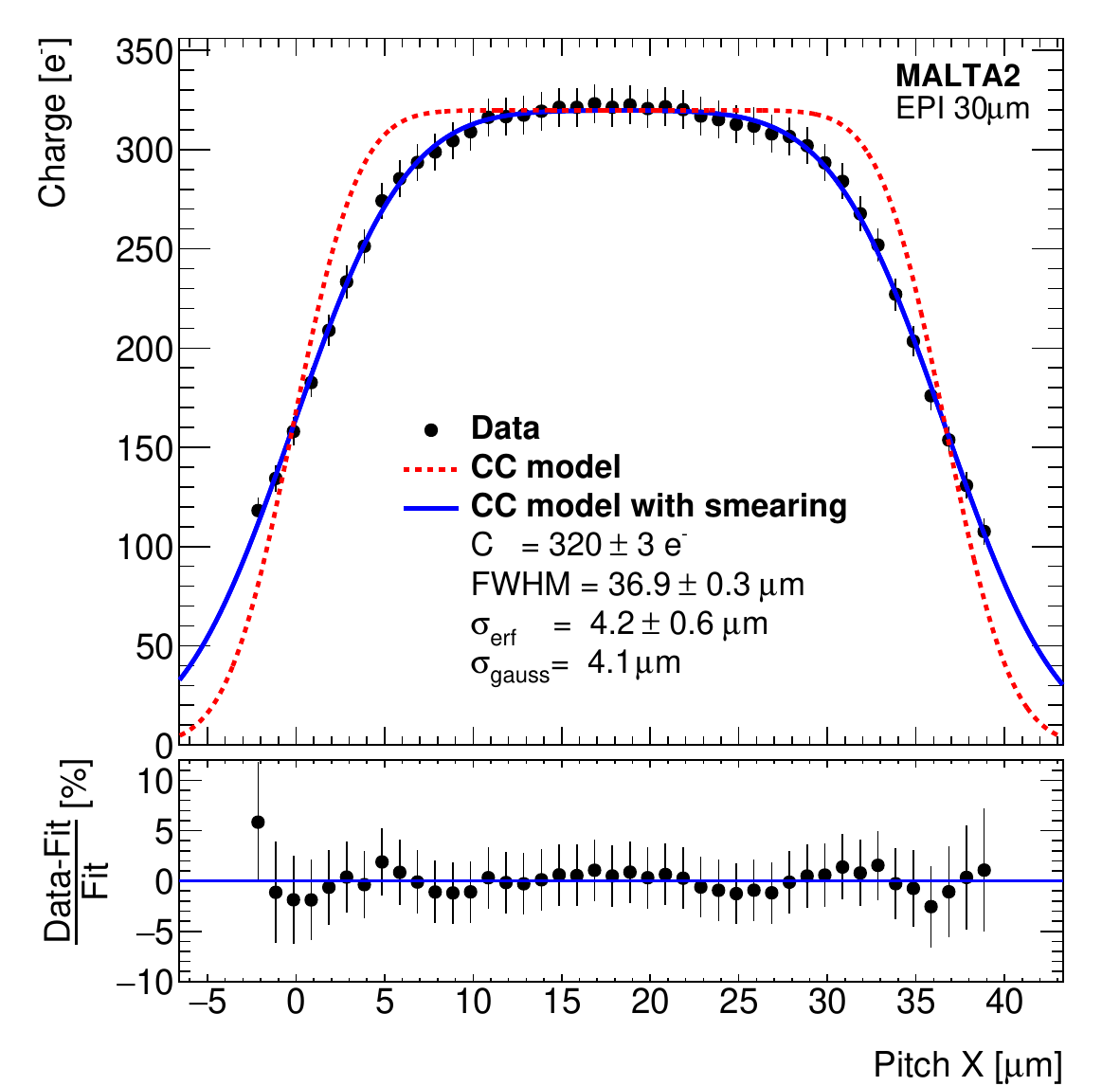}
\caption{1D scan along the X-dimension.}\label{fig:DigitalTCT_1DPitch}
\end{subfigure}
\caption{
(a) quantifies the area inside a contour at different thresholds.
It matches the expectation at half maximum.
The scan in (b) goes through the pixel center.
The charge collection model (red dashed line) describes the data when convolved with the laser beam width $\sigma_{\mathrm{gauss}}$ (blue line).
The error bars represent the uncertainty from charge calibration.% of about \SI{3}{\percent}.
}\label{fig:Analysis}
\end{figure}

The 2D charge map of the central pixel is used to quantify its sensitive area, defined as the area in which the charge exceeds a given threshold (\cref{fig:SensitiveArea}).
The area decreases continuously with threshold and vanishes once the threshold reaches the maximum charge. 
At half maximum (red dotted line), it agrees with the expectation from a fully depleted epitaxial layer (gray dotted line).
At lower thresholds, the area exceeds the nominal pixel area due to charge sharing from neighboring pixels.
This threshold dependence is relevant for grazing-angle studies, which reconstruct a depletion depth from the cluster size.
The reconstructed depth decreases with threshold, as observed in ref.~\cite{Berlea_TCT}, because the sensitive area decreases. %through cluster size measurements at varying angles
The following section derives a charge collection model from the measurement.
%The depth reconstruction is sensitive to the geometric size of the

\section{Charge collection model}\label{sec:1DCCModel}
To parameterize the charge collection, a pixel is scanned in \SI{1}{\micro\meter} steps along the X-dimension through its center $\left(Z=\SI{15}{\micro\meter}\right)$, as shown in \cref{fig:DigitalTCT_1DPitch}.
The collected charge is modeled as a function of the position $x$ relative to the pixel by
\begin{equation}
\label{eq:CC}
\mathrm{CC}(x;C,\sigma_{\mathrm{erf}},\mathrm{FWHM}) = \frac{C}{2}\left[\mathrm{erf}\left(\frac{x}{\sigma_{\mathrm{erf}}}\right)-\mathrm{erf}\left(\frac{x-\mathrm{FWHM}}{\sigma_{\mathrm{erf}}}\right)\right].
\end{equation}
$C$ is a normalization constant and $\sigma_{\mathrm{erf}}$ the width of the error-function at the pixel edges.
The distance between the two edges is the full width at half maximum (FWHM).
The charge collection model (red dashed line) is convolved with a Gaussian distribution, which accounts for the smearing due to the width of the laser beam $\sigma_{\mathrm{gauss}} = \SI{4.1}{\micro\meter}$.
%The width is determined by scanning a sharp edge and finding the mini 
The convolution (blue line) is fitted to the data.
All three model parameters are left free in the fit.
The deviation between data and fit is well contained within the 3\% uncertainty of charge calibration~\cite{MALTA2Calibration}.
%\subsection{Pitch axis scan}
The fit describes the data well at both edges and yields $\sigma_{\mathrm{erf}} = \SI{4.2\pm0.6}{\micro\meter}$.
The FWHM of $ \SI{36.9\pm0.3}{\micro\meter}$ is close to the nominal pixel pitch of \SI{36.4}{\micro\meter}.
To test whether this pixel is representative, the same fit is performed for more than one hundred pixels. % matrix of $17\times7$
The results are compatible, with a mean and standard deviation of $\sigma_{\mathrm{erf}} =\SI{4.4\pm0.7}{\micro\meter}$.
Both values agree with an independent charge collection measurement using minimum-ionizing particles at the CERN SPS, which yielded a width of $\SI{4.3\pm0.3}{\micro\meter}$~\cite{MALTA2Param}.
Thus, the laser and test beam measurements are consistently modeled by \cref{eq:CC}, which serves as input to the parametric simulation.

\begin{figure} 
\centering
\begin{subfigure}[t]{0.59\textwidth} 
\centering
\includegraphics[width=\linewidth]{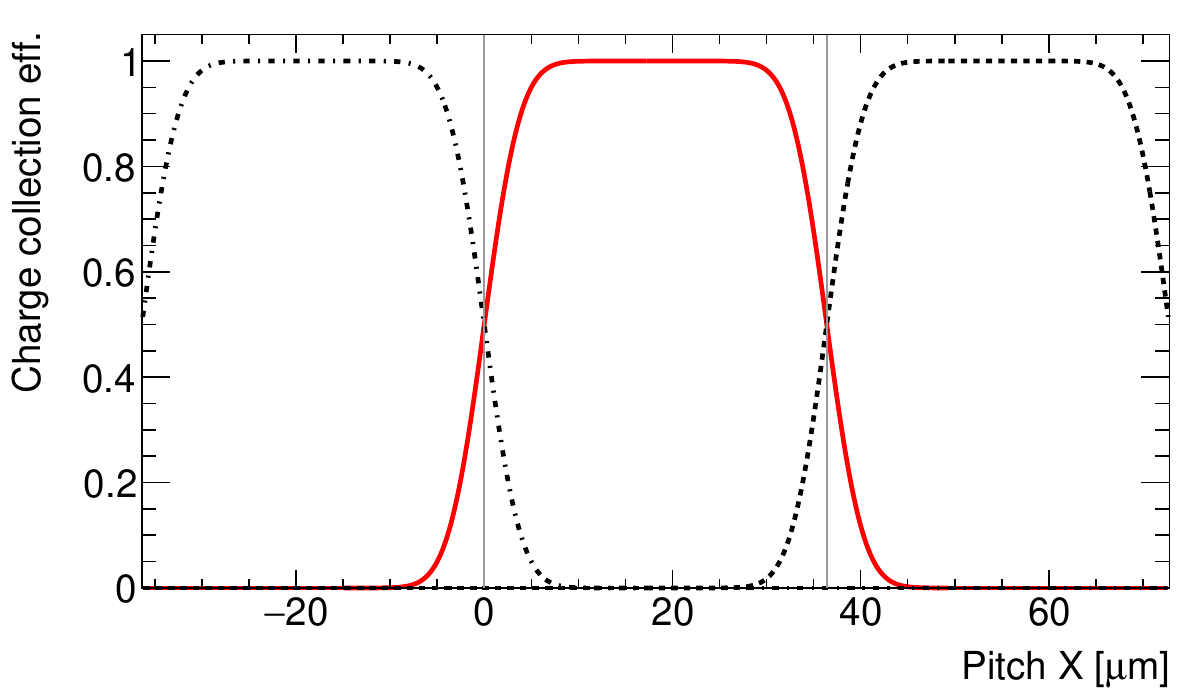}
\caption{1D charge collection model}\label{fig:1DIllustration}
\end{subfigure}
\hfill
\begin{subfigure}[t]{0.39\textwidth}
\centering
\includegraphics[width=\linewidth]{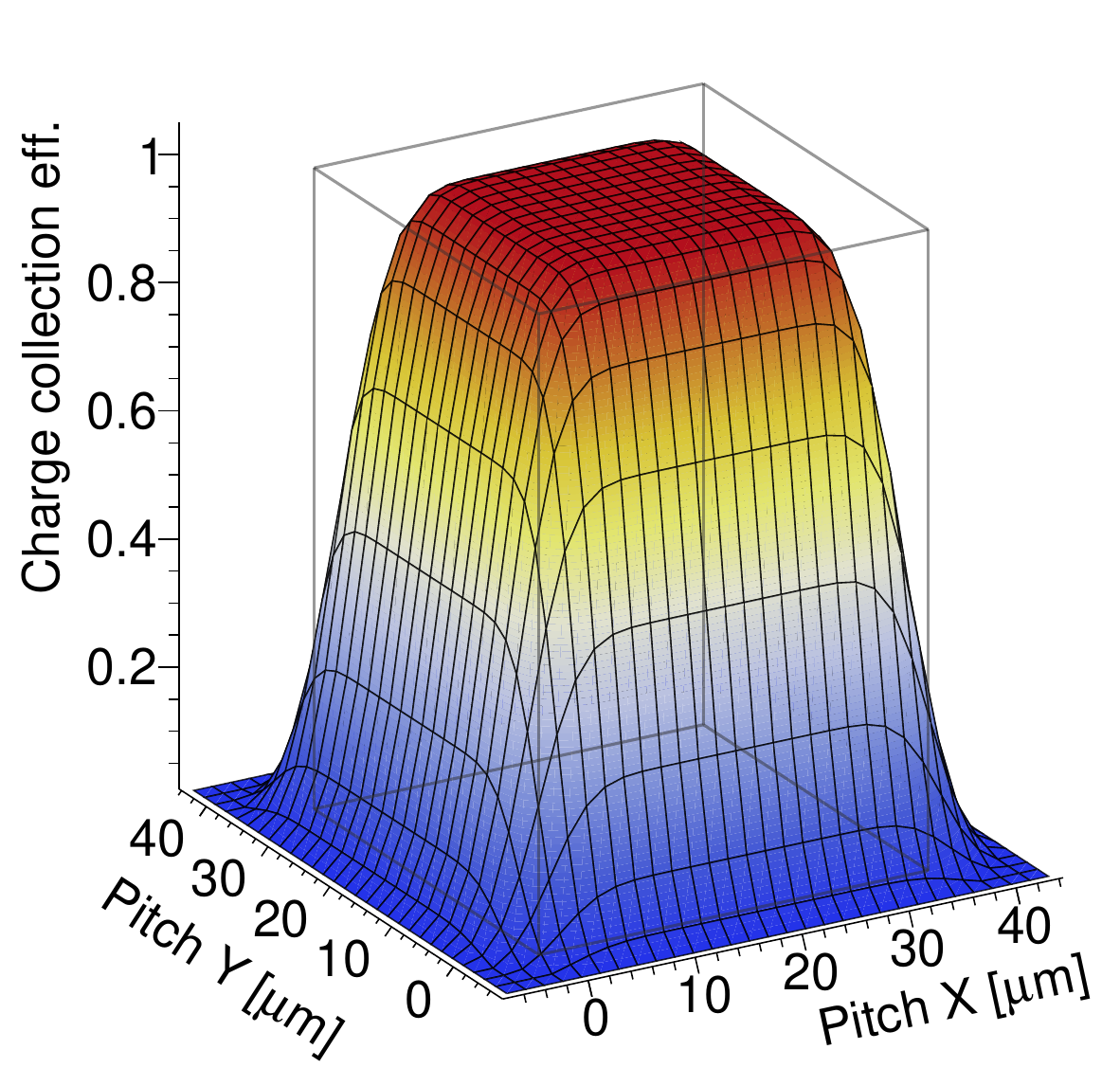}
\caption{2D model}\label{fig:2DIllustration}
\end{subfigure}
\caption{Charge collection efficiency along the pitch dimension. 
(a) shows the model of a seed pixel and its two neighbors in the X-dimension.
It is defined by \cref{eq:CC}.
(b) illustrates the symmetric 2D model for the seed pixel.
}\label{fig:Illustrations}
\end{figure}

\section{Parametric simulation}\label{sec:ParSim}
%A realistic sensor simulation is constructed from the presented charge collection model.
The charge collection model is the main input to the simulation~\cite{FastSim}.
It distributes a charge deposition in Geant4~\cite{Geant4} among the seed pixel and its neighbors according to \cref{eq:CC}, evaluated as $\mathrm{CC}(x;C=1,\sigma_{\mathrm{erf}}=\SI{4.3}{\micro\meter},\mathrm{FWHM}=\SI{36.4}{\micro\meter})$.
\Cref{fig:1DIllustration} illustrates these fractions for the seed pixel (red line) and its neighbors (dashed lines).
Near the pixel boundaries (vertical gray lines), up to 50\% of the charge is shared with a neighbor.
The same model is applied in the Y-dimension, which leads to the 2D charge collection in \cref{fig:2DIllustration}.
In addition to charge sharing, the simulation incorporates the time walk of the front-end and the digital readout logic.

\begin{figure} 
\centering
\begin{subfigure}[t]{0.49\textwidth} 
\centering
\includegraphics[trim=15 10 0 0,clip,width=\linewidth]{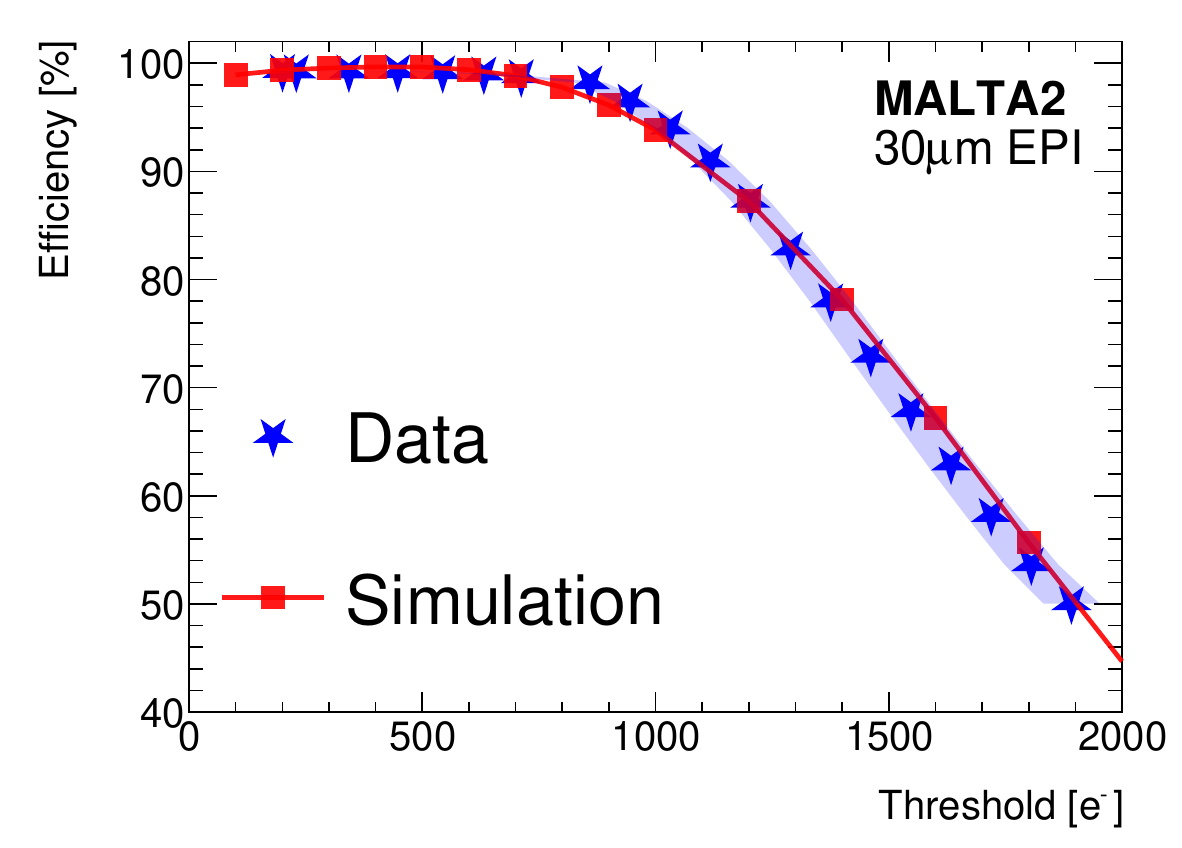}
\caption{Average efficiency}\label{fig:Eff_vs_Thr}
\end{subfigure}
\hfill
\begin{subfigure}[t]{0.49\textwidth}
\centering
\includegraphics[trim=15 10 0 0,clip, width=\linewidth]{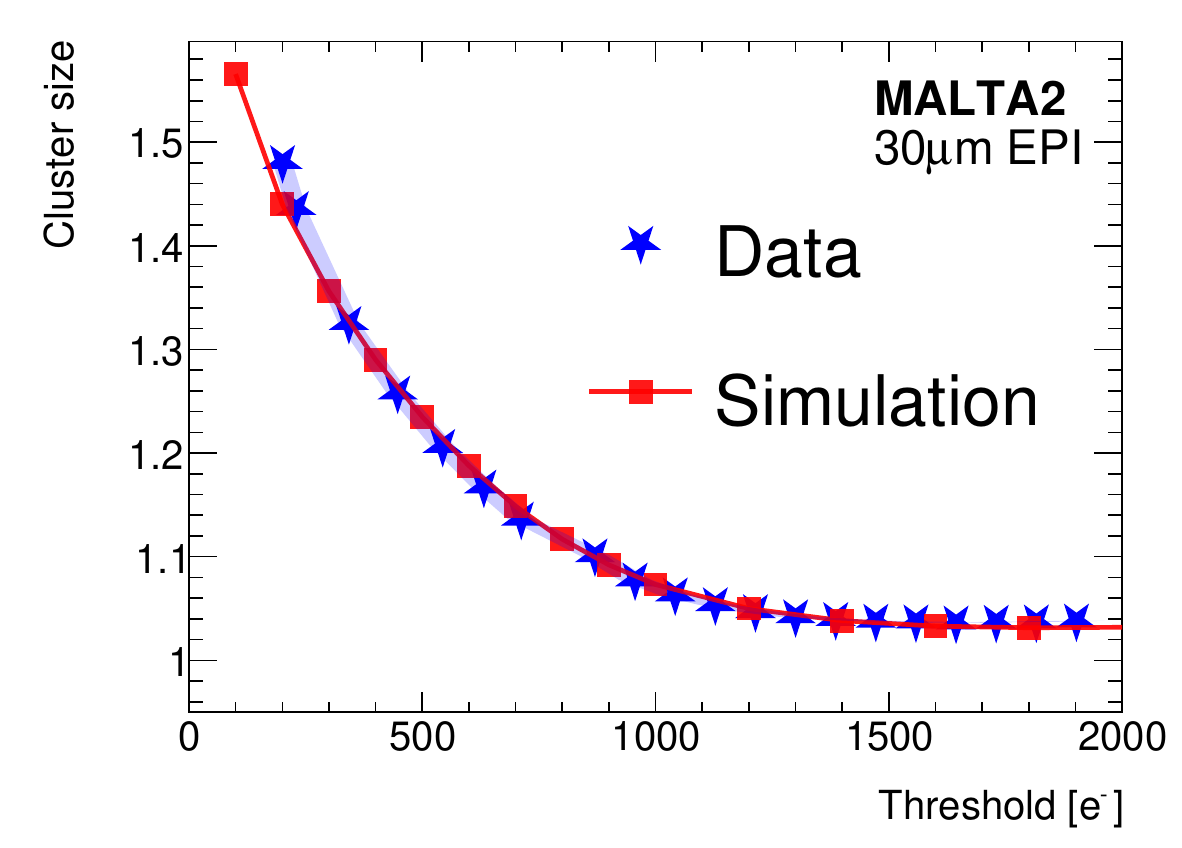}
\caption{Average cluster size}\label{fig:Cl_vs_Thr}
\end{subfigure}
\caption{Data and simulation are compared for the efficiency (a) and cluster size (b) of a MALTA2 sensor.
The uncertainty band represents the threshold uncertainty of 3\%.
}\label{fig:Eff_Cl}
\end{figure}

The simulation is validated with test beam data measured at the CERN SPS.
Charged hadrons of \SI[per-mode = symbol]{120}{\giga\electronvolt\per\clight} are incident on the front of the sensor.
While the typical operational threshold is $200\,\mathrm{e}^-$ it can be raised up to $2000\,\mathrm{e}^-$.
\Cref{fig:Eff_Cl} compares efficiency and cluster size as a function of threshold.
Simulation and data agree within the threshold uncertainty, which is represented by the uncertainty band.
Additionally, the simulation reproduces the measured in-pixel profiles of efficiency and cluster size, as discussed in detail in ref.~\cite{FastSim}.
This validates the accuracy of the simulation across the full threshold range.

\section{Conclusion}
Digital Edge-TCT is a practical tool for characterizing multiple pixels of a MAPS.
Since it makes use of the standard digital readout, it does not require dedicated analog test pixels or an oscilloscope.
%Despite the digital hit data, the charge collection can nevertheless be measured and modeled.
For MALTA2, the charge collection is reconstructed from binary hit data and modeled by error-functions with a width of about \SI{4}{\micro\meter} at the pixel edges. %, in agreement with test beam data.
This model is the input to a parametric simulation, which reproduces the measured efficiency and cluster size over a wide threshold range.
The simulation is currently used to investigate the application of MALTA2 in larger detector geometries and to optimize the readout for a new sensor design.
%The improvement of hit rato improve the detection at hit rate capability in a new sensor design.
%This data-driven simulation approach is pursued without the input of doping profiles or other proprietary information.

\acknowledgments
This project has received funding from the European Union's Horizon 2020 Research and Innovation programme under Grant Agreement numbers: 101004761 (AIDAinnova), 675587 (STREAM), 654168 (AIDA-2020).

\end{document}